\documentclass[notitlepage,nofootinbib,preprintnumbers,amssymb]{revtex4-1}

\usepackage{amsfonts,amssymb,amsmath,graphicx,color,bm}
\definecolor{ultramarine}{rgb}{0.07, 0.04, 0.56}
\definecolor{cadmiumgreen}{rgb}{0.0, 0.42, 0.24}
\definecolor{indigo(dye)}{rgb}{0.0, 0.25, 0.42}
\usepackage[linktocpage=true]{hyperref}
\hypersetup{
colorlinks=true,
citecolor=ultramarine,
linkcolor=cadmiumgreen,
urlcolor=indigo(dye),
}

\newcommand{\f}[2]{\frac{#1}{#2}}  
\newcommand{\mk}[1]{\left( #1 \right)}  
\newcommand{\kk}[1]{\left[ #1 \right]}  
\newcommand{\ck}[1]{\left\{ #1 \right\}}  
\newcommand{\be}{\begin{equation}}  
\newcommand{\ee}{\end{equation}}
\newcommand{\Mpl}{M_{\rm Pl}}
\newcommand{\e}{\epsilon}

\begin{document}

\preprint{RESCEU-51/14}  

\title{  
Inflation with a constant rate of roll
}

\author{Hayato Motohashi$^{~1}$}
\author{Alexei A.\ Starobinsky$^{~2,3}$}
\author{Jun'ichi Yokoyama$^{~2,4,5}$}

\address{
$^{1}$ Kavli Institute for Cosmological Physics,
The University of Chicago, Chicago, Illinois 60637, U.S.A. \\
$^{2}$ Research Center for the Early Universe (RESCEU),
Graduate School of Science, The University of Tokyo, Tokyo 113-0033, Japan \\
$^{3}$ L. D. Landau Institute for Theoretical Physics RAS,
Moscow 119334, Russia \\
$^{4}$ Department of Physics, Graduate School of Science, The University of Tokyo, Tokyo 113-0033, Japan \\
$^{5}$ Kavli Institute for the Physics and Mathematics of the Universe
(Kavli IPMU), WPI, UTIAS,
The University of Tokyo, Kashiwa, Chiba 277-8568, Japan
}

\begin{abstract}
We consider an inflationary scenario where the rate of inflaton roll 
defined by $\ddot\phi/H\dot \phi$ remains constant. The rate of roll is 
small for slow-roll inflation, while a generic rate of roll leads to the 
interesting case of `constant-roll' inflation. We find a general exact 
solution for the inflaton potential required for such inflaton behaviour. 
In this model, due to non-slow evolution of background, the would-be 
decaying mode of linear scalar (curvature) perturbations may not be 
neglected. It can even grow for some values of the model parameter, 
while the other mode always remains constant. However, this always occurs 
for unstable solutions which are not attractors for the given potential.
The most interesting particular cases of constant-roll inflation remaining viable 
with the most recent observational data are quadratic hilltop inflation (with 
cutoff) and natural inflation (with an additional negative cosmological constant). 
In these cases even-order slow-roll parameters approach non-negligible 
constants while the odd ones are asymptotically vanishing in the 
quasi-de Sitter regime.
\end{abstract}


\maketitle  


\section{Introduction}

The inflationary paradigm based on the assumption of the existence of 
a quasi-de Sitter stage in the early Universe before the hot radiation 
dominated Big Bang \cite{Starobinsky:1980te,Sato:1980yn,Guth:1980zm,Linde:1981mu,Albrecht:1982wi} 
has now become a well established part of modern cosmology. It makes 
definite predictions about present properties of the Universe which 
have been confirmed by numerous observations. The most common way to 
realize this quasi-de Sitter expansion is to employ a scalar field 
(inflaton) with an approximately flat potential and consider a 
slow-roll solution of the field equations. It is also possible to 
realize inflationary expansion without introducing such an inflaton 
field by upgrading the gravitational action from the Einstein-Hilbert 
one to a nontrivial function of the Ricci scalar, $f(R)$. However, 
using the conformal transformation, equations for an $f(R)$ type 
inflation can be transformed to the Einstein-Hilbert action with a 
canonical scalar field dubbed scalaron with its potential defined by 
the form of $f(R)$. In this approach, the $R^2$ inflation \cite{Starobinsky:1980te}, 
including its generalization to accommodate 
present dark energy \cite{Appleby:2009uf,Motohashi:2012tt,Nishizawa:2014zra} 
or to consider a 
small deviation from it \cite{DeFelice:2010aj,Martin:2014vha,Motohashi:2014tra}, 
can generate  
primordial scalar and tensor metric fluctuations of the same form as 
scalar field inflation.

Given that slow-roll inflationary models with an approximately flat 
inflaton potential yield a simple realization of inflation with viable 
observational predictions, it is natural to ask what happens if we omit 
the assumption of the inflaton slow roll. The slow-roll approximation 
assumes that in the Klein-Gordon equation for the inflaton given by 
\be \ddot\phi+3H\dot\phi+\f{\partial V}{\partial \phi}=0, \ee
the second derivative term $\ddot \phi$ is negligible compared to other 
terms. The necessity to go beyond this approximation and to use some
exact solutions instead has been already occurred in a number of cases, 
in particular, when the inflaton potential $V(\phi)$ has some local 
non-analytic feature \cite{Starobinsky:1992ts} or is near its local extremum \cite{Inoue:2001zt}, 
if $\partial V/\partial \phi=0$ holds for an extended period 
\cite{Tsamis:2003px,Kinney:2005vj,Namjoo:2012aa} (the latter case 
dubbed the `ultra-slow-roll' inflation) and if the period of fast-roll 
inflation preceded about 50 e-folds of slow-roll inflation 
\cite{Contaldi:2003zv,Lello:2013awa,Hazra:2014jka}. An important new feature appearing
in all these cases is that the curvature power spectrum becomes evolving 
on super-Hubble scales temporarily. Furthermore, the non-Gaussianity 
consistency relation for single field inflation can be violated \cite{Namjoo:2012aa}.

Motivated by these phenomenologically interesting features, the 
ultra-slow-roll inflation was generalized in \cite{Martin:2012pe}.
Starting from an assumption of a constant rate of roll with 
$\ddot\phi/H\dot\phi=-3-\alpha$, where nonzero $\alpha$ implies deviation 
from the flat potential, they derived a potential that satisfies the 
assumption approximately. However, it is actually possible to construct 
an exact solution under this assumption, which is the main topic of the 
present paper. We refer to this class of models as the `constant-roll' 
inflation. We shall show that the general exact solution for these models 
includes power-law inflation~\cite{Abbott:1984fp,Lucchin:1984yf} and 
other two solutions: One of them is equivalent to a solution previously found in 
\cite{Barrow:1994nt} in a different context, 
and the other one amounts to a particular case of 
hilltop inflation~\cite{Boubekeur:2005zm} with cutoff
or natural inflation~\cite{Freese:1990rb} with an additional negative cosmological constant. 
We shall also investigate the evolution of scalar (curvature) 
perturbations in these models.
In general, these constant-roll models may have super-Hubble evolution of the
scalar perturbation, which is the same situation as the ultra-slow-roll inflation.
However, we shall show that for the particular cases of hilltop inflation 
and natural inflation do not suffer from it and 
they actually remain observationally feasible 
with the most recent constraints for the tilt of the scalar power spectrum and 
the tensor-to-scalar ratio.

The rest of the paper is organized as follows.
In \S\ref{sec-back}, we determine the inflaton potential required for constant-roll inflation. 
We show that it is possible to derive exact solution 
for inflationary dynamics without assumptions made in the literature. 
We shall see that in order to make the inflationary regime an attractor, 
one needs the constant-roll parameter $\alpha<-3/2$.
In \S\ref{sec-pert}, we explore scalar and tensor perturbations generated during 
constant-roll inflation. 
In \S\ref{ssec-scalar}, we calculate power spectrum of the curvature perturbation 
and show that $\alpha \gtrsim 0$ or $\alpha \lesssim -3$ provides a slightly red-tilted spectrum, 
the latter of which has an attractor inflationary regime.
In \S\ref{ssec-superH}, we confirm that the super-Hubble evolution of the curvature perturbations 
in non-attractor model $\alpha>-3/2$, whereas $\alpha<-3/2$ serves a constant mode and 
a decaying mode on super-Hubble scales as the standard slow-roll inflation.
Therefore, $\alpha \lesssim -3$ is observationally viable and 
analytically solvable constant-roll inflation model, 
which includes particular cases of hilltop inflation and natural inflation. 
In \S\ref{ssec-tensor}, we addressed tensor perturbation.
We then conclude in \S\ref{sec-conc}. 
Throughout the paper, we will work in the natural unit where $c=1$, 
and the metric signature is $(-+++)$.

\section{Constant-roll inflation}
\label{sec-back}

We consider inflation driven by a minimally coupled scalar field defined by the 
action
\be S=\int d^4x \sqrt{-g} \kk{\f{\Mpl^2}{2}R -\f{1}{2}g^{\mu\nu}
\partial_\mu\phi\partial_\nu \phi-V(\phi)}, \ee
where $\Mpl= (8\pi G)^{-1/2}$.
Working in the flat Friedmann-Lema\^{i}tre-Robertson-Walker (FLRW) metric, the 
Friedmann equation and the equation of motion for the scalar field are given by
\begin{align}
\label{eeq1} 3\Mpl^2H^2&=\f{\dot \phi^2}{2}+V, \\
\label{eeq2} -2\Mpl^2\dot H&=\dot \phi^2,\\
\label{eomp} \ddot\phi+3H\dot\phi+\f{\partial V}{\partial \phi}&=0,
\end{align} 
where $a$ is the scale factor, a dot denotes the derivative with respect to $t$, 
$H\equiv \dot a/a$ is the Hubble parameter.
Inflationary evolution is characterized by the slow roll parameters:
\be \e_1\equiv -\f{\dot H}{H^2},\quad \e_{n+1}\equiv \f{\dot \e_n}{H\e_n}. \ee

In the slow-roll inflation, the slow-roll parameters are small, $|\e_n| \ll 1$, 
and the first terms of the right-hand side of \eqref{eeq1} and the left-hand side 
of \eqref{eomp} can be neglected. As a result, an approximately flat spectrum 
of scalar (curvature) perturbations can be obtained, whose exact form depends on 
the functional shape of the potential $V(\phi)$ during the quasi-de Sitter 
expansion regime. Here, instead, we are interested in a different regime when 
$\ddot\phi$ is not negligible in \eqref{eomp}. Following \cite{Martin:2012pe}, 
we adopt an ansatz of a constant rate of roll $\ddot\phi/H\dot\phi$ during
inflation and parametrize it as
\be \label{cond} \ddot\phi=-(3+\alpha)H\dot \phi, \ee
with an arbitrary constant value of $\alpha$. The standard slow-roll inflation 
occurs if $\alpha\simeq -3$ while the `ultra-slow-roll' case corresponds to 
$\alpha=0$. Thus, the constant-roll inflation interpolates between these two 
regimes. By the way, this shows that the term `ultra-slow-roll inflation' is 
rather misleading. In fact, this regime has to be considered as a specific case
of fast-roll inflation since the slow-roll approximation breaks down during it.

In \cite{Martin:2012pe}, by neglecting $\dot\phi^2/2$ term in \eqref{eeq1}, an 
inflaton potential which satisfies the ansatz \eqref{cond} was derived. However, 
in this paper we shall show that it is actually possible to construct an inflaton 
potential which realizes the relation \eqref{cond} without any approximation and 
to investigate its dynamics fully analytically. In order to solve the system of 
equations \eqref{eeq1}--\eqref{eomp} with the condition \eqref{cond}, we consider 
the Hubble parameter as a function of the inflaton field $H=H(\phi)$ and use 
the Hamiltonian-Jacobi-like formalism \cite{Muslimov:1990be,Salopek:1990jq}. This approach 
is applicable as long as $t=t(\phi)$ is a single-valued function, i.e.\ 
$\dot\phi\not= 0$. Since $\dot H=\dot \phi dH/d\phi$ holds in this approach, we 
can rewrite \eqref{eeq2} as
\be \label{dphi} \dot \phi = -2\Mpl^2\f{dH}{d\phi}. \ee
Plugging the time derivative of \eqref{dphi} to \eqref{cond}, we obtain the 
differential equation for the Hubble parameter with respect to $\phi$,
\be \f{d^2H}{d\phi^2}=\f{3+\alpha}{2\Mpl^2}H. \ee
The general solution of this equation is 
\be \label{gensol}
H(\phi)=C_1\exp\mk{ \sqrt{\f{3+\alpha}{2}}\f{\phi}{\Mpl} }+
C_2\exp\mk{ -{\sqrt{\f{3+\alpha}{2}}\f{\phi}{\Mpl}} }. 
\ee
For $\alpha<-3$, the exponents become imaginary, and then $C_2=C_1^{\ast}$. 
Using \eqref{eeq1} and \eqref{dphi}, we find the inflaton potential required 
for the exact solution \eqref{gensol}: 
\begin{align} \label{Veq}
V(\phi)&=\Mpl^2\kk{3H^2-2\Mpl^2\mk{\f{dH}{d\phi}}^2} \notag\\
&=\Mpl^2\kk{-\alpha C_1^2\exp\mk{\sqrt{2(3+\alpha)}\f{\phi}{\Mpl}}
+2(6+\alpha)C_1C_2
-\alpha C_2^2\exp\mk{-\sqrt{2(3+\alpha)}\f{\phi}{\Mpl}}}.
\end{align} 
Then we can derive evolution of inflaton $\phi(t)$ by solving \eqref{dphi}, 
and obtain $H(t)$ by plugging $\phi(t)$ back into \eqref{gensol}. However, it
should be kept in mind that the Hamilton-Jacobi formalism produces one solution
$\phi(t)$ for the derived potential $V(\phi)$ only which needs not be an 
attractor solution (even an intermediate one). So, its stability with respect 
to all FLRW solutions with the same potential has to be checked each time.

The most interesting particular solutions for $\alpha>-3$ are
\begin{align}
\label{Hsol1} H&=M e^{\pm \sqrt{\f{3+\alpha}{2}}\f{\phi}{\Mpl}},\\
\label{Hsol2} H&=M \cosh \mk{\sqrt{\f{3+\alpha}{2}}\f{\phi}{\Mpl}}, \\
\label{Hsol3} H&=M \sinh \mk{\sqrt{\f{3+\alpha}{2}}\f{\phi}{\Mpl}},
\end{align}
where $M$ is an integration constant which determines the amplitude of the 
power spectrum of the curvature perturbation. Note that the third solution
describes a bounce. For $\alpha<-3$, only the last two solutions have physical
sense, and the hyperbolic functions in them have to be substituted by the
corresponding trigonometric ones.  
Consequently, for $\alpha<-3$ two solutions are the same up to a field redefinition.

Let us begin with the first solution \eqref{Hsol1}. Since two solutions are 
equivalent under $\phi\to -\phi$, we focus on 
$H=M e^{\sqrt{\f{3+\alpha}{2}}\f{\phi}{\Mpl}}$. From \eqref{Veq}, we then obtain
\begin{align}
V(\phi)&=- \alpha M^2\Mpl^2 \exp\mk{\sqrt{2(3+\alpha)}\f{\phi}{\Mpl}}, \label{V-1}\\
\phi&=-\sqrt{\f{2}{3+\alpha}}\Mpl\ln[(3+\alpha)Mt], \\
H&=\f{1}{(3+\alpha)t},\\
a&\propto t^{\frac{1}{3+\alpha}}.
\end{align}
The positivity of $V(\phi)$ requires $\alpha<0$, and for $-2<\alpha<-3$ this is 
nothing but the power-law inflation~\cite{Abbott:1984fp,Lucchin:1984yf}. As is 
well known, it leads to the constant slopes $n_s-1=n_t=2(3+2\alpha)/(2+\alpha)<0$ 
of the power spectra of primordial scalar and tensor perturbations. Of course, to 
obtain an exit from inflation, one has to assume that the potential \eqref{V-1} 
changes its form and quickly approaches zero at some value of $\phi$. But from 
the observational point of view, power-law inflation is certainly excluded because 
of the absence of the large amount of primordial tensor perturbations 
(gravitational waves) that would be $r=8(1-n_s)\approx 0.28$ in this case.

Plugging the second solution \eqref{Hsol2} to \eqref{Veq}, we obtain for 
$\alpha>-3$
\begin{align}
\label{Vsol2} V(\phi)&=3M^2\Mpl^2\kk{ 1+\f{\alpha}{6}
\ck{1-\cosh\mk{\sqrt{2(3+\alpha)}\f{\phi}{\Mpl}}} },\\
\label{phisol2} \phi&=\Mpl \sqrt{\f{2}{3+\alpha}} 
\ln\kk{\coth\mk{\f{3+\alpha}{2}Mt}},\\
H&= M \coth [ (3 + \alpha) M t], \\
\label{sf-2} a &\propto \sinh^{1/(3+\alpha)} \kk{(3+\alpha)Mt}.
\end{align}
This solution is equivalent to a solution found in \cite{Barrow:1994nt} in a 
different context. Note also that \eqref{sf-2} with $\alpha=3(w-1)/2$ 
mimics the evolution of the FLRW model with a cosmological constant filled by 
an ideal fluid with the equation of state $p=w\rho$ with $w={\rm const}$. 
This case encompasses a more special case of mimicking the $\Lambda$CDM model with $w=0$,
which is found in \cite{Motohashi:2014una}. 
For $-3<\alpha<0$,
$V(\phi)$ has a minimum at $\phi=0$. Thus, to end inflation, a kind of phase 
transition at this point has to be assumed additionally, e.g.\ similar to that 
occurs in hybrid inflation \cite{Linde:1993cn}.

Contrary, for $\alpha<-3$ with \eqref{Hsol2} we obtain
\begin{align}
\label{Vsol2-} V(\phi)&= 3M^2\Mpl^2\kk{1+\f{\alpha}{6}\ck{1-\cos \mk{\sqrt{2|3+\alpha|} \f{\phi}{\Mpl} } } } , \\
\phi&=  2\sqrt{\f{2}{|3+\alpha|}}\Mpl {\rm arctan} (e^{|3+\alpha|Mt}) , \label{phisol2-}\\
H&= -M\tanh \mk{|3+\alpha|Mt} , \\
a&\propto \cosh^{-1/|3+\alpha|} \mk{|3+\alpha|Mt}, \label{asol2-}
\end{align}
which is a particular case of hilltop inflation~\cite{Boubekeur:2005zm} 
with the inverted quadratic potential near the origin. 
As usual in this case, we assume that $\phi\to +0$ at $Mt\to -\infty$, 
so the Hubble parameter remains positive and approaches $M$. 
Then inflation takes place as the inflaton rolls down from the origin, 
and ends near the point where the potential crosses zero. 
In fact, the latter has to be changed somewhere before this point and 
the initial condition for $\phi$ may be made less restrictive.  
We shall also see that this case includes natural inflation~\cite{Freese:1990rb} 
with an additional negative cosmological constant.  
We discuss the above points in \S\ref{ssec-superH}.

Finally, the third solution \eqref{Hsol3} yields 
\begin{align}
V(\phi)&=-3M^2\Mpl^2\kk{ 1+\f{\alpha}{6}\ck{1+\cosh\mk{\sqrt{2(3+\alpha)}\f{\phi}{\Mpl}}} },\\
\phi&=-2\Mpl \sqrt{\f{2}{3+\alpha}}{\rm arctanh}\kk{\tan\mk{\f{3+\alpha}{2}Mt}},\\
H&=-M\tan[(3+\alpha)Mt],\\
a &\propto \cos^{1/(3+\alpha)} [(3+\alpha)Mt].
\end{align}
Although this is a mathematically allowed solution, it has $\ddot a(t)<0$. Therefore, it 
cannot describe an inflationary model in the usual sense.

\begin{figure}[t]
	\centering
	\includegraphics[width=75mm]{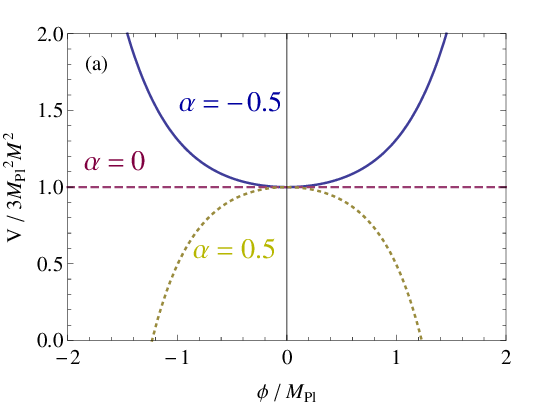}
	\includegraphics[width=75mm]{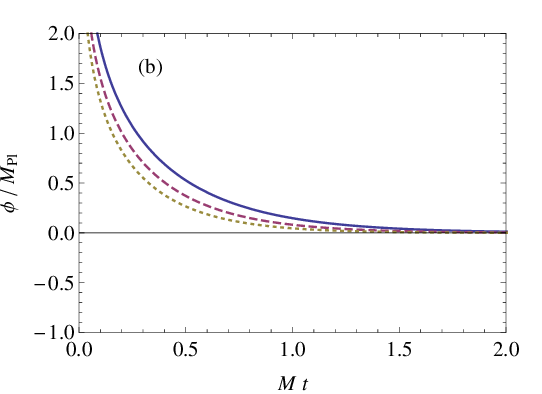}
	\includegraphics[width=75mm]{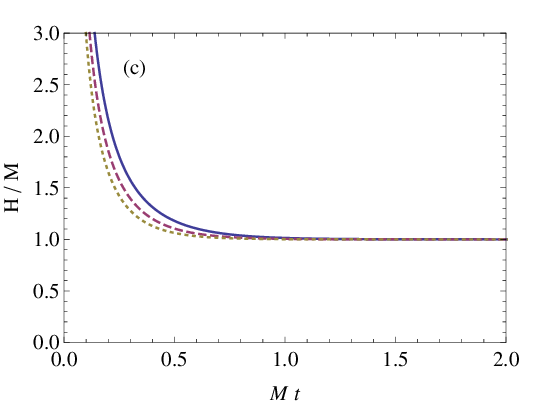}
	\includegraphics[width=75mm]{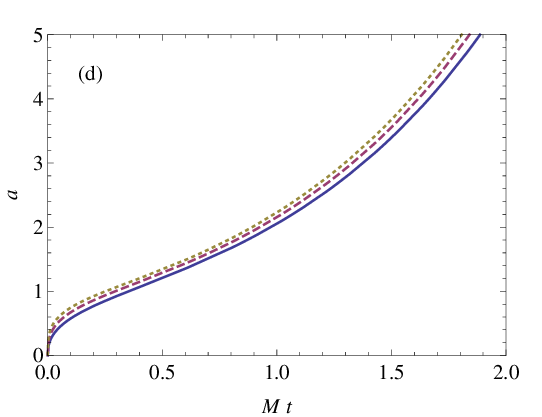}
	\includegraphics[width=75mm]{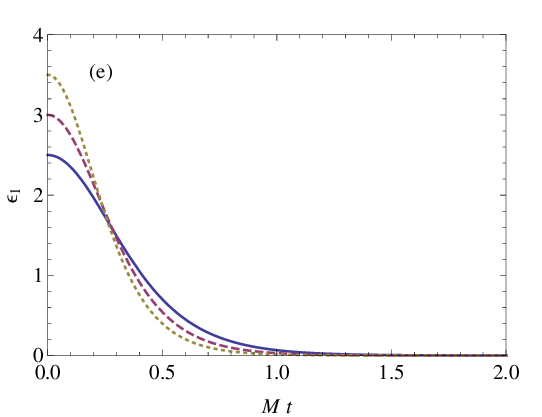}
	\includegraphics[width=75mm]{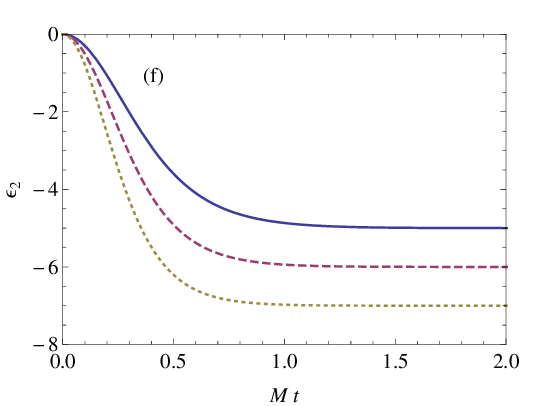}
	\caption{(a) Potential \eqref{Vsol2}, and its exact solution for (b) inflaton $\phi$, (c) Hubble parameter $H$, (d) scale factor $a$, and slow-roll parameters (e) $\e_1\equiv -\dot H/H^2$, (f) $\e_2\equiv \dot \e_1/H\e_1$, for  $\alpha=-0.5$ (blue solid), $0$ (red dashed), $0.5$ (yellow dotted). While $\e_2$ approaches to $-2(3+\alpha)$, which is not necessarily negligible, the Hubble parameter approaches constant and the scale factor evolves as exponentially. The higher slow-roll parameters are given by $\e_{2n+1}=2\e_1$ and $\e_{2n}=\e_2$. 
	}
	\label{fig:infdyn}
\end{figure}

For the following, we investigate the second solution \eqref{Hsol2}, 
i.e.\ \eqref{Vsol2}--\eqref{sf-2} with $\alpha>-3$ and 
\eqref{Vsol2-}--\eqref{asol2-} with $\alpha<-3$, 
and clarify if they serve observationally feasible inflationary scenario.
In particular, we shall check the potential and the analytic solutions 
for the former case \eqref{Vsol2} below  
as the potential for the latter case \eqref{Vsol2-} amounts to a particular case 
of natural inflation or hilltop inflation with $V\propto {\rm const}-\phi^2$ around $\phi=0$.
We present the potential \eqref{Vsol2} in Fig.~\ref{fig:infdyn}~(a) 
for $\alpha=-0.5$ (blue solid), $0$ (red dashed), $0.5$ (yellow dotted). 
Clearly, $\alpha=0$ yields a flat potential and it reduces 
to the case of the original ultra-slow-roll inflation.
The evolution of inflaton is depicted in Fig.~\ref{fig:infdyn}~(b).
The field value monotonically decreases and approaches to the origin.
For the case of $\alpha=0.5$, the inflaton climbs up the potential 
and approaches to the top of the potential at the origin. 
We note that the Hubble parameter in Fig.~\ref{fig:infdyn}~(c) 
approaches constant as time goes by, 
which leads to de Sitter expansion of the universe, i.e.\
$a(t) \propto e^{Mt}$ for $Mt\gg 1$, 
which is displayed in Fig.~\ref{fig:infdyn}~(d). 
In this limit, the effective inflaton mass squared becomes 
$m^2(\phi)\equiv \partial^2V/\partial \phi^2=-\alpha(3+\alpha)M^2$. 
Then the generic solution for the inflaton is the
sum of two exponents $e^{\alpha Mt}$ and $e^{-(3+\alpha)Mt}$.
However, our constant-roll solution \eqref{phisol2} chooses 
the second exponent only due to the definition \eqref{cond}. 
The same situation happens for $\alpha<-3$ in the limit of $Mt\to -\infty$.
This is an illustration of the remark above that 
the Hamilton-Jacobi method of finding exact solutions 
produces only one solution for the potential $V(\phi)$ 
determined by its use at the same time.

We can also express the conformal time $\tau=\int dt/a$ 
analytically in terms of the Gauss' hypergeometric function ${}_2F_1$. 
For \eqref{Vsol2} with $\alpha>-3$, we obtain 
\be 
\tau=\f{(-1)^{\f{4+\alpha}{2(3+\alpha)}}}{M(3+\alpha)} 
\kk{ \cosh [(3+\alpha)Mt] 
{}_2F_1 \mk{\f{1}{2},\f{4+\alpha}{2(3+\alpha)},\f{3}{2};\cosh^2 [(3+\alpha)Mt]}
- (-1)^{-1/2} {}_2F_1 \mk{\f{1}{2},\f{5+2\alpha}{2(3+\alpha)},\f{3}{2};1} },
\ee
where we fixed the integration constant to make $\tau\to -0$ as $Mt\to \infty$.
On the other hand, for \eqref{Vsol2-} with $\alpha<-3$, we obtain
\begin{align} 
\tau = \f{i}{(2+\alpha)M} \kk{ \cosh^{ \f{2+\alpha}{3+\alpha} }[(3+\alpha)Mt] 
{}_2F_1\mk{\f{1}{2}, \f{2+\alpha}{2(3+\alpha)}, \f{8+3\alpha}{2(3+\alpha)}; \cosh^2[(3+\alpha)Mt] }
- \f{\sqrt{\pi} \Gamma \mk{ \f{8+3\alpha}{2(3+\alpha)} } }{\Gamma \mk{ \f{5+2\alpha}{2(3+\alpha)}  }} } , 
\end{align}
where the integration constant is determined by $\tau\to -0$ as $Mt\to -0$.
Of course, to get a realistic model, we have to cut the potential somewhere before it becomes negative. 
In that case the end of the inflation should be defined accordingly and we would choose integration constant to set $\tau\to -0$ at the end of inflation.

The above derivation has been done fully analytically. 
Let us check that the slow-roll approximation indeed breaks down in these models.
It is easy to show that the second slow-roll parameter 
is non-negligible by plugging its definition 
$\e_2=2\e_1+2\ddot \phi/(H\dot\phi)$ and the condition \eqref{cond}.
This fact has been pointed out from the beginning of this class of models \cite{Tsamis:2003px}.
Our exact solution allows us to see the violation in higher order slow-roll parameters.
For \eqref{Vsol2} with $\alpha>-3$,
the first slow-roll parameter $\e_1$ is given by
\be \label{eps1} \e_1=\f{3+\alpha}{\cosh^2 [(3+\alpha)Mt]}=\f{3+\alpha}{a^{2(3+\alpha)}+1}, \ee
and for higher order slow-roll parameters $\e_n$ with $n\geq 1$ are given by 
\be \label{eps2} \e_{2n}=-2(3+\alpha)\tanh^2 [(3+\alpha)Mt], 
\quad \e_{2n+1}=\f{2(3+\alpha)}{\cosh^2 [(3+\alpha)Mt]}. \ee
We show their evolution in Fig.~\ref{fig:infdyn}~(e) and (f). In particular, for $Mt\gg 1$,
\be \label{symsp} 2\e_1= \e_{2n+1} \simeq 2(3+\alpha) a^{-2(3+\alpha)}, \quad \e_{2n} \simeq -2(3+\alpha). \ee
Thus, the odd-order slow-roll parameters asymptotically approach to zero, 
while the even ones approach to $-2(3+\alpha)$, which is 
not necessarily negligible, rather, $\e_{2n}$ can be of order unity. 
Thus the slow-roll approximation clearly breaks down.
Nevertheless, the Hubble parameter approaches to a constant 
$H\simeq M$ and the scale factor grows as exponentially, so this is still inflation.
Likewise, for \eqref{Vsol2-} with $\alpha<-3$,
the slow-roll parameters are given by
\be
2\epsilon_1=\epsilon_{2n+1}=-\f{2(3+\alpha)}{\sinh^2[(3+\alpha)Mt]}, \quad
\epsilon_{2n}=-\f{2(3+\alpha)}{\tanh^2[(3+\alpha)Mt]},
\ee
and their asymptotic behaviors are $2\epsilon_1=\epsilon_{2n+1} \to 0$ and 
$\epsilon_{2n}\to -2(3+\alpha)$ as $Mt\to -\infty$, which are the same with the case for $\alpha>-3$.

\begin{figure}[t]
	\centering
	\includegraphics[width=75mm]{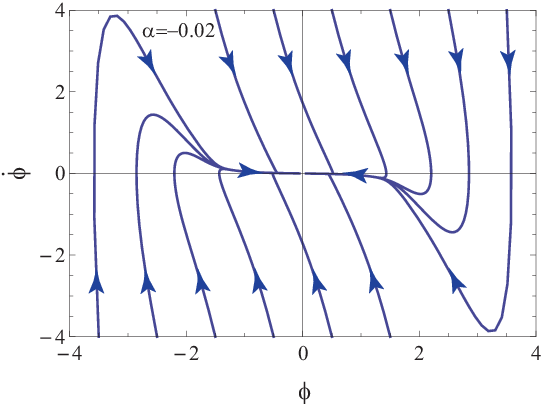}
	\includegraphics[width=75mm]{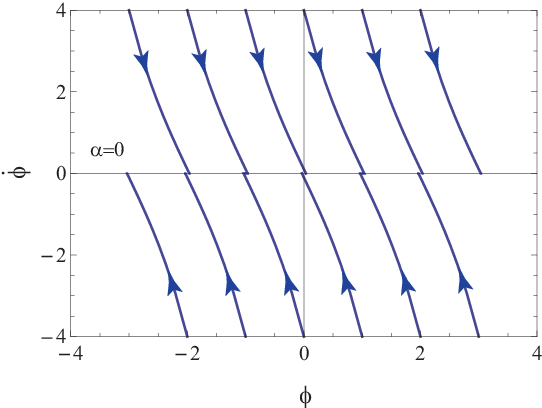}
	\includegraphics[width=75mm]{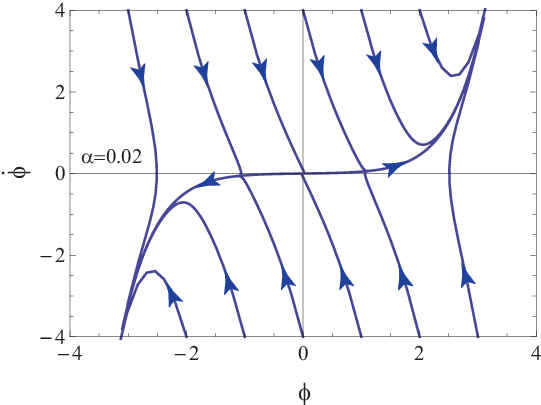}
	\caption{Phase space diagram for $\alpha=-0.02$ (top left), $\alpha=0$ (top right) and $0.02$ (bottom).
	}
	\label{fig:phase}
\end{figure}

We proceed to check whether our exact solution is an attractor solution or not.  
We numerically solved inflationary dynamics under the 
assumption \eqref{cond} with various initial conditions 
and obtained the phase space diagram as depicted in Fig.~\ref{fig:phase} for different $\alpha$.
The top left panel of Fig.~\ref{fig:phase} for $\alpha=-0.02$ 
shows a typical attractor behavior where the phase space flow converges to $\dot\phi=0$ and $\phi=0$.
This implies that the inflaton approaches to the global minimum of the potential at $\phi=0$. 
Thus the analytic solution with negative $\alpha$ is an attractor. 
The top right panel for $\alpha=0$ of Fig.~\ref{fig:phase} shows phase space 
flow converges various points depending on initial conditions.
This is not surprising because we note that for $\alpha=0$ the potential 
is exactly constant $V=3M^2\Mpl^2$, so it is invariant under translation 
$\phi\to \phi+{\rm const}$.
The scalar field moves on the flat potential with the Hubble friction, and 
it eventually stops at a point which depends on an initial condition.
In this case, we need to recover the integration constant for our analytic 
solution \eqref{phisol2} and determine it according to an initial condition. 
Finally, the bottom panel of Fig.~\ref{fig:phase} for $\alpha=0.02$ shows 
that the inflaton goes to $\dot \phi\to \pm \infty$ and $\phi\to \pm \infty$, 
respectively. Therefore the analytic solution for positive $\alpha$ is not an attractor. 
This is obvious because the potential is not bounded from below. The scalar 
field rolls down to $\phi\to\pm\infty$ since $V\to -\infty$. Hence, we need 
to make fine-tuning of the initial condition for scalar field obeying the 
analytical solution \eqref{phisol2} for positive $\alpha$.

For a general $\alpha>-3$, this special constant-roll solution for which
$\phi(t)\propto \exp\kk{-(3+\alpha)Mt}$ in the limit $Mt\gg 1$ is stable, i.e.\
it is an attractor during expansion, if it grows faster with time than the 
second linearly independent solution for the same potential which is 
$\phi(t)\propto \exp \mk{\alpha Mt}$. Therefore, the stability condition is 
$-(3+\alpha)>\alpha$, or $\alpha<-3/2$ that includes the slow-roll case
$\alpha\approx -3$. The same formally refers to the $\alpha<-3$ case which
is stable as far as inflation goes on.

\section{Evolution of scalar and tensor perturbation}
\label{sec-pert}

\subsection{Scalar perturbation}
\label{ssec-scalar}

We consider the gauge invariant curvature perturbation 
$\zeta_k$ in our models \eqref{Vsol2} and \eqref{Vsol2-}.
It relates to the metric perturbation through 
$\delta g_{ij} = a^2(1-2\zeta)\delta_{ij}$ in a gauge $\delta\phi=0$.
The evolution of the mode function $v_k\equiv \sqrt{2}\Mpl z \zeta_k$ with 
$z\equiv a \sqrt{\e_1}$ is governed by the Mukhanov-Sasaki equation~\cite{Mukhanov:1985rz,Sasaki:1986hm}:
\be \label{mseq} v''_k+\mk{k^2-\f{z''}{z}}v_k=0, \ee
where a prime denotes a derivative with respect to the conformal time $\tau$.
The potential term $z''/z$ is exactly expressed in terms of slow-roll parameters:
\be \f{z''}{z}=a^2H^2\mk{2-\e_1+\f{3}{2}\e_2+\f{1}{4}\e_2^2-\f{1}{2}\e_1\e_2+\f{1}{2}\e_2\e_3}. \ee
We can solve \eqref{mseq} by the standard treatment of the slow-roll inflation 
except that $\e_{2n}$ is not negligible whereas $\e_{2n-1}$ is. Starting from the 
sub-Hubble regime where $k^2\gg z''/z$, \eqref{mseq} reduces to $v''_k+k^2v_k=0$.
We choose the adiabatic vacuum initial condition, i.e.\ no particles and minimum energy 
at $\tau\to-\infty$:
\be \label{BDvac} v_k(\tau)= \f{e^{-ik\tau}}{\sqrt{2k}}. \ee
As the mode $k$ approaches to the Hubble radius crossing, the potential term 
$z''/z$ becomes non-negligible. 
Since during generation of the curvature perturbation 
$2\epsilon_1=\epsilon_{2n+1} \to 0$ and $\epsilon_{2n}\to -2(3+\alpha)$
for both cases with $\alpha>-3$ and $\alpha<-3$, 
we can simplify the potential term as 
\be \f{z''}{z}\simeq \f{(1+\alpha)(2+\alpha)}{\tau^2}
=\f{\nu^2-1/4}{\tau^2}, \ee
where
\be \nu\equiv \sqrt{(1+\alpha)(2+\alpha)+\f{1}{4}}=\left| \alpha+\f{3}{2} \right|. \ee
With \eqref{BDvac} as the boundary condition, the solution is given 
in terms of Hankel function as usual,
\be v_k(\tau)=\f{\sqrt{-\pi\tau}}{2} H_\nu^{(1)}(-k\tau). \ee
The power spectrum of the curvature perturbation is then given by
\be \label{Ds2} \Delta^2_s(k)\equiv \f{k^3}{2\pi^2}|\zeta_k|^2
=\f{H^2}{8\pi^2\Mpl^2\e_1} \mk{\f{k}{aH}}^3 \f{\pi}{2} |H_\nu^{(1)}(-k\tau)|^2. \ee
Using the asymptotic formula $\lim_{x\to 0} H_\nu^{(1)}(x) 
\simeq -\f{i}{\pi}\Gamma(\nu) \mk{\f{x}{2}}^{-\nu}$, we obtain
\be \Delta^2_s(k)= \f{H^2}{8\pi^2\Mpl^2\e_1} \f{2^{2\nu-1}[\Gamma(\nu)]^2}{\pi} \mk{\f{k}{aH}}^{3-2\nu}, \ee
and the spectral index is then $n_s-1=3-2\nu$.
For given $n_s$ from observations, the constant-roll parameter $\alpha$ is given by
\be \alpha= \f{1-n_s}{2} \quad \text{or} \quad \f{n_s-7}{2}. \ee
For instance, for $n_s=0.96$, we can choose $\alpha=0.02$ or $-3.02$.
However, as we discussed in the previous section, the analytic solution 
for $\alpha=0.02$ is not an attractor solution. In addition, we should be careful 
for a possible  evolution of the curvature perturbation on super-Hubble scales 
\cite{Martin:2012pe}. Indeed, from \eqref{Ds2} we confirmed that 
$\Delta^2_s(k)\propto H^{-1+2\nu}\e_1^{-1}a^{-3+2\nu} \propto a^{2\alpha+3+|2\alpha+3|}$ evolves for $\alpha=0.02$, but 
approaches to a constant for $\alpha=-3.02$. 
Therefore $\alpha=-3.02$ is viable, and 
the standard slow-roll approximation works in this case. However, our exact solution
gives a possibility to sum all higher-order corrections to it, at least for the 
background evolution.

\subsection{Super-Hubble evolution}
\label{ssec-superH}

Alternatively, we can examine the behavior of curvature perturbations in the 
super-Hubble regime $k^2\ll z''/z$ by directly solving $v''_k-(z''/z)v_k=0$. 
The formal solution for this equation is given by a linear combination of 
$z$ and $z \int d\tau /z^2$.
Since $\zeta_k\propto v_k/z$, we thus obtain
\be \label{zsuh} \zeta_k = A_k + B_k  \int \f{dt}{a^3\e_1}, \ee
where $A_k$ and $B_k$ are integration constants.
The first term expresses the constant mode of the
curvature perturbations. For slow-roll inflation, the second term 
yields decaying mode, thus only first term remains, and as a result the curvature 
perturbation is conserved outside the Hubble radius. 
However, this is not necessarily the case for constant-roll inflation, as we shall see below.

For the model \eqref{Vsol2} with $\alpha>-3$, 
we can perform the integral of the second term of \eqref{zsuh} analytically 
using \eqref{sf-2} and \eqref{eps1}. 
The result is 
\begin{align}
\int \f{dt}{a^3\e_1}&= 2\Mpl^2\int \f{H^2}{a^3\dot\phi^2}\, dt 
\notag\\
&\propto \int dt \f{\cosh^2 [(3+\alpha)Mt] \sinh^{-\f{3}{3+\alpha}} [(3+\alpha)Mt]}{3+\alpha}  \notag\\
&=\f{(-1)^{\f{6+\alpha}{2(3+\alpha)}}}{3M(3+\alpha)^2}
\cosh^3 [(3+\alpha)Mt]~ u \mk{\f{3}{2},\f{6+\alpha}{2(3+\alpha)},\f{5}{2};\cosh^2 [(3+\alpha)Mt]}, 
\label{vinteg}
\end{align}
where $u(a,b,c;x)$ is a solution for the hypergeometric differential equation:
\be x(1-x)\f{d^2u}{dx^2}+[c-(a+b+1)x]\f{du}{dx}-ab u=0. \ee 
It has two independent solutions and the solutions that converge for $x\gg 1$ are 
expressed in terms of the hypergeometric function.
(i) For ${\rm Re}(a+b-c)>0$, two solutions are 
\begin{align}
u_1&=(-x)^{b-c}(1-x)^{c-a-b} {}_2F_1(1-b,c-b,a-b+1;1/x) \simeq x^{-a}, \\
u_2&=(-x)^{a-c}(1-x)^{c-a-b} {}_2F_1(1-a,c-a,b-a+1;1/x) \simeq x^{-b},
\end{align}
and (ii) for ${\rm Re}(a+b-c)<0$,
\begin{align} 
u_3 &=(-x)^{a} {}_2F_1(a,a-c+1,a-b+1;1/x) \simeq x^{-a}, \\
u_4&=(-x)^{b} {}_2F_1(b,b-c+1,b-a+1;1/x) \simeq x^{-b}.
\end{align}
For the case of \eqref{vinteg}, $-3<\alpha<0$ amounts to the case (i), and the integral is evaluated by
$x^{3/2}u_1\simeq {\rm const}$ and $x^{3/2}u_2\simeq x^{\f{3+2\alpha}{2(3+\alpha)}}$,
where $x=\cosh^2 [(3+\alpha)Mt]$. 
Therefore, the first solution gives a constant mode, 
which is absorbed into the definition of $A_k$, whereas the second 
solution is a decaying mode for $-3<\alpha<-3/2$,
or a growing mode for $-3/2<\alpha<0$. 
The remaining region $\alpha>0$ amounts to the case (ii), 
and two solutions behave as $x^{3/2}u_3\simeq {\rm const}$ and 
$x^{3/2}u_4\simeq x^{\f{3+2\alpha}{2(3+\alpha)}}$.
These asymptotic behavior of two solutions are the same with the case (i). 
The first solution gives a constant mode while the second one 
is a growing mode as $\alpha>0$.

On the other hand,
for the model \eqref{Vsol2-} with $\alpha<-3$, 
we can prove that the second term of \eqref{zsuh} yields sum of constant mode and decaying mode as $Mt$ increases from $-\infty$. 
The integral is given by 
\be \int \f{dt}{a^3\epsilon_1} \propto -\f{i}{\alpha(3+\alpha) M} \cosh^{\alpha/(3+\alpha)}[(3+\alpha)Mt]~ 
u \mk{ -\f{1}{2},\f{\alpha}{2(3+\alpha)},\f{3(2+\alpha)}{2(3+\alpha)}; \cosh^2[(3+\alpha)Mt] }. \ee
As $a+b-c=-3/2<0$ for the arguments of $u(a,b,c;x)$, its two independent solutions are given by $u_3$ and $u_4$. 
Thus the asymptotic behavior of the above integral is 
$x^{b}u_3 \simeq x^{b-a}=x^{\f{3+2\alpha}{2(3+\alpha)}}=x^{1+\f{3}{|3+\alpha|}}$ and 
$x^{b}u_4 \simeq {\rm const}$. 
The latter mode is a constant mode, and the former mode is a decaying mode as $Mt$ increases from $-\infty$, 
namely, as $x=\cosh^2 [(3+\alpha)Mt]$ decreases from $+\infty$.

In summary, in the super-Hubble regime
a curvature perturbation has two modes which asymptotically approach to
\be {\rm const}, \quad {\rm and} \quad \cosh^{\f{3+2\alpha}{3+\alpha}} [(3+\alpha)Mt], \ee 
respectively.
The latter one is a decaying mode for $\alpha <-3/2$, but it is a growing mode for $\alpha >-3/2$. 
The condition for the decaying mode coincides with the attractor condition for background evolution derived in the previous section.

In comparison with slow-roll inflation, where we always have a constant mode 
and a decaying mode outside the Hubble radius, we may have a growing mode in 
constant-roll inflation. This is because $\e_1$ decays as $a^{-2(3+\alpha)}$, 
which is faster than $a^{-3}$ for $\alpha>-3/2$. As a result, 
$\zeta\simeq \int dt/(a^3\e_1)$ grows in this case.  
Any model with background evolution with $\e_2 = d\ln \e_1/d\ln a < -3$ possesses 
the super-Hubble evolution of the curvature perturbation.
This situation is similar to what happens in 
the chaotic new inflation model where curvature perturbation grows anomalously in
between chaotic and new inflation stages \cite{Saito:2008em}. However, as was 
shown at the end of the previous section, $\alpha>-3/2$ is just the condition that 
our particular constant-roll solution for the given potential is {\em not} an 
attractor during expansion. Thus, in this case such solution can typically occur 
for no more than a few e-folds.

\begin{figure}[t]
	\centering
	\includegraphics[width=75mm]{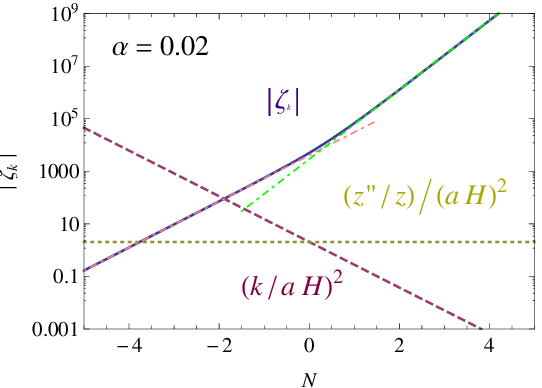}
	\includegraphics[width=75mm]{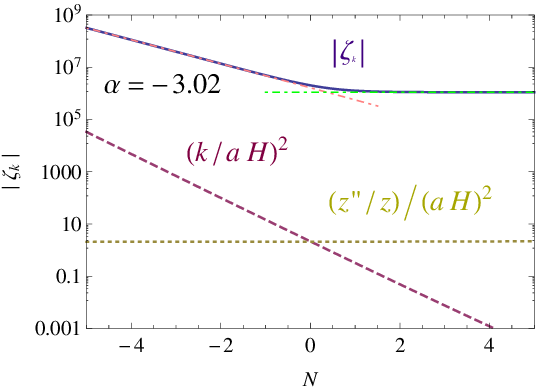}
	\caption{Evolution of the amplitude of the curvature perturbation (blue solid) 
	for $\alpha=0.02$ (left) and $-3.02$ (right) as a function of e-folds $N=\ln a$ obtained by integrating \eqref{mseq} numerically. The analytic solutions for $(k/aH)^2$ (red dashed), and $(z''/z)/(aH)^2$ (yellow dotted) are also presented. Two dotted-dashed lines are asymptotic analytic solution for the curvature perturbation, which are proportional to $(a\sqrt{\e_1})^{-1}\propto \sinh^{\f{1}{3+\alpha}} [(3+\alpha)Mt] /\cosh [(3+\alpha)Mt]$ for the earlier phase (pink dotted-dashed) and $\cosh^{\f{3+2\alpha}{3+\alpha}} [(3+\alpha)Mt]$ for the latter phase (green dotted-dashed), respectively.}
	\label{fig:zeta}
\end{figure}

To confirm our discussion above, we numerically integrated \eqref{mseq} with analytic 
solution for the background quantities derived in the previous section. The result is shown 
in Fig.~\ref{fig:zeta}. 
In the left and right panel, we present the result for \eqref{Vsol2} with $\alpha=0.02$, 
and the result for \eqref{Vsol2-} with $\alpha=-3.02$, respectively. 
The solid blue line represents the amplitude of the curvature perturbation $|\zeta_k|$, 
the red dashed line expresses $(k/aH)^2$ and yellow dotted line is $(z''/z)/(aH)^2$. 
We normalized the e-folds $N=\ln a$ so that $k^2=z''/z$ at $N=0$, 
which is about a half e-fold earlier than the Hubble radius crossing $k=aH$. 
Two dotted-dashed lines are asymptotic analytic solution for the curvature perturbation. The 
pink dotted-dashed line is the one for the sub-Hubble evolution, which is proportional to 
$(a\sqrt{\e_1})^{-1}\propto \sinh^{\f{1}{3+\alpha}} [(3+\alpha)Mt] /\cosh [(3+\alpha)Mt]$. 
The green dotted-dashed line is the one for the super-Hubble evolution, which is 
proportional to $\cosh^{\f{3+2\alpha}{3+\alpha}} [(3+\alpha)Mt]$.
In the left panel, for $\alpha=0.02$ we see that the curvature perturbation switches its evolution 
at the Hubble radius crossing, and continuously evolves even on the super-Hubble regime. 
On the other hand, as depicted in the right panel, for $\alpha=-3.02$
the curvature perturbation remains constant after the Hubble radius exit, as expected.

The situation for $\alpha=0.02$ above is similar to the generalized ultra-slow-roll 
inflation~\cite{Martin:2012pe}, which also has a growing mode and results in an 
unwanted amplification of the curvature perturbation outside the Hubble radius. Combined 
with the observational value of the amplitude of the scalar power spectrum, this
growth of the curvature perturbation should be compensated by imposing extremely 
small energy scale of inflation, at least $M\simeq 10^{-42}\Mpl$. This is much 
smaller than the BBN bound $M>{\cal O} ({\rm MeV})$. The same occurs for constant-roll 
inflation. Thus, the case $\alpha \gtrsim 0$ with \eqref{Vsol2} does not give us 
a possibility to construct a physically relevant inflationary model.

Much more perspective appears the case $\alpha \lesssim -3$ with \eqref{Vsol2-}. 
However, in this case the potential 
has to be modified at some value $\phi=\phi_0$ which is less than the critical value 
$\phi_c=\pi\Mpl/\sqrt{2|3+\alpha|}$ where $V(\phi)<0$ and $H$ changes the sign. 
The aim of this modification is to finish inflation at $\phi=\phi_0$. 
This situation is completely analogous to what happens in the case of power-law inflation. 
So, then our solution can be applied not to the whole evolution of the Universe but to 
its evolution during inflation  only (apart from its very end) including the observable range of e-folds. 
The best-fit value of the constant-roll parameter $\alpha$ and 
the prediction for the tensor-to-scalar ratio $r$ will depend on $\phi_0$. We shall see below 
that  $|3+\alpha|$ has to be small for any reasonable value of $\phi_0$, so $\phi_c\gg \Mpl$
anyway. Thus, depending on the value of $\phi_0$, such a model can describe both small-field 
and large-field inflation.

Let us first consider the case when $\phi_0\ll \phi_c$. Then $V(\phi)$ can be approximated as 
\be \label{top-quad}
V(\phi)=3M^2\Mpl^2 \left(1-\frac {\alpha (3+\alpha)\phi^2}{6\Mpl^2}\right),
\ee
during the whole inflation. In this case $n_s-1$ is constant and equal to $-2\alpha (3+\alpha)/3$. 
So, the best fit from the recent CMB data is $\alpha \approx -3.02$.  
As was shown above, the curvature perturbation remains constant on super-Hubble 
scales in this case. The number of e-folds from the end of inflation is
\be
N=\Mpl^{-2}\int_{\phi}^{\phi_0}\frac{V}{dV/d\phi}d\phi =\frac{3}{\alpha (3+\alpha)}
\ln \left(\frac{\phi_0}{\phi}\right)\approx 50 \ln \left(\frac{\phi_0}{\phi}\right).
\ee
Thus, due to numerical coincidence between the measured value of $1-n_s$ and $2/N_{H_0}$
where $N_{H_0}=50-60$ corresponds to the present Hubble radius $H_0^{-1}$, 
the inflaton field value in the observable range of e-folds during inflation is $\approx\phi_0e^{-1}$.  
This provides a possibility to significantly enlarge possible range for an initial condition for the inflaton 
field at the local beginning of inflation: it is sufficient to have its any value less than $\approx\phi_0e^{-1}$.

If now $\phi_0\lesssim \phi_c$, but not specifically close to it, the potential \eqref{Vsol2-} and the 
corresponding exact constant-roll solution \eqref{phisol2-}--\eqref{asol2-} may be taken as they are up 
to $\phi=\phi_0$. Then the model looks like the so called `natural inflation'~\cite{Freese:1990rb} 
but with the additional negative cosmological constant $\Lambda=M^2(3+\alpha)<0$. 
Also, it is always large-field inflation. In this case, the measured value of $n_s-1$ 
does not determine $\alpha$ unambiguously, it provides only an 
upper bound on $|3+\alpha|$. Still $|3+\alpha|$ has to be small, so $|\Lambda| \ll M^2$. As a result, this 
cosmological constant is subdominant during inflation. However, it affects higher-order slow-roll corrections. 
Therefore, the model \eqref{Vsol2-} represents a novel example of viable and exactly analytically solvable 
(for the evolution of background and perturbations in the super-Hubble regime) model of inflation.

The limiting, although somewhat fine-tuned case, occurs if $\phi_0=\phi_c-{\cal O} ( \Mpl )$.  Then
at the last stage of inflation, when $\phi\gg \Mpl$ and $\tilde\phi\gg \Mpl$ where $\tilde\phi =
\phi_c - \phi$, we get
\be \label{quad}
H\approx M\sqrt{\frac{|3+\alpha|}{2}}\frac{\tilde\phi}{\Mpl},~~V\approx \frac{\alpha (3+\alpha)}{2}
M^2\tilde\phi^2~,
\ee
i.e.\ just quadratic chaotic inflation. Then $n_s-1=-2/N$, and we obtain its correct value
independently of the value of $\alpha$. However, the upper limit on $\alpha$ follows from the condition 
that the approximation \eqref{quad} still works for $N=N_{H_0}$ when $\tilde\phi\approx 15\Mpl$. 
From the condition that the latter quantity should be $\ll \phi_c$, it follows that $|3+\alpha|\ll 0.02$, 
i.e.\ much less than in the hilltop case $\phi_0\ll \phi_c$ considered above.

\subsection{Tensor perturbation}
\label{ssec-tensor}

Finally, we consider the tensor perturbation $\delta g_{ij}=a^2h_{ij}$ 
for our models. The evolution equation for the mode function of the tensor perturbation
$u_{k,\lambda}\equiv a \Mpl h_{k,\lambda}/2$ is given by 
\be u_{k,\lambda}''+ \mk{k^2-\f{a''}{a}}u_{k,\lambda}=0, \ee
where $\lambda=+,\times$ denotes the two polarization modes of the gravitational waves.
As $a''/a=(aH)^2(2-\epsilon_1)\simeq (2+3\epsilon_1)/\tau^2\simeq 2/\tau^2$, we obtain 
(in agreement with \cite{Starobinsky:1979ty})
\be \Delta^2_t = \f{2H^2}{\pi^2\Mpl^2}, \ee
and thus the standard consistency relation for the tensor-to-scalar ratio
\be r\equiv \f{\Delta^2_t}{\Delta^2_s}\approx 16\epsilon_1 = 32\Mpl^2\left(\frac{d\ln H(\phi)}{d\phi}\right)^2
\approx 8\Mpl^2\left(\frac{d\ln V(\phi)}{d\phi}\right)^2\ee
holds with both $\approx$ signs becoming $=$ in the leading order of the slow-roll approximation.

For the quadratic hilltop case $\phi_0\ll \phi_c$ we, therefore, get $r=8(3+\alpha)^2 \phi^2/\Mpl^2$. So,
parametrically by powers of $|3+\alpha|$, $r$ is of the order of $N^{-2}$. However, actually $r$ can
be much less than the latter quantity if $\phi_0\ll \Mpl$. To get $r\sim N^{-2}$ as, e.g.\ in the
$R+R^2$ inflationary model~\cite{Starobinsky:1980te}, one needs $\phi_0\sim \Mpl$. Larger values of $r$
in constant-roll inflation, of the order $N^{-1}$ parametrically, 
can be obtained if $\Mpl \ll \phi_0\sim \phi_c$. 
Then the exact background solution \eqref{phisol2-}--\eqref{asol2-} has to be used. 
Finally, in the limiting case 
$\phi_0=\phi_c-{\cal O} ( \Mpl )$, $r$ reaches the value $8/N$ ($\sim 0.14$ for $N=N_{H_0}$) as for the 
quadratic chaotic inflation, but this model lies just beyond the $2\sigma$ CL contour for the recent Planck 
data~\cite{Planck:2013jfk}.

\section{Conclusion}
\label{sec-conc}

We have investigated an inflationary scenario where the rate of roll defined by 
$\ddot\phi/ H\dot \phi=-3-\alpha$ remains constant. This class includes slow-roll inflation 
with negligible rate of roll and fast-roll inflation, in particular, the so called
`ultra-slow-roll' one. We find all exact solutions satisfying the constant-rate-of-roll 
ansatz. They include power-law inflation, the solution found in~\cite{Barrow:1994nt} 
in a different context, and particular (and somewhat modified) cases of hilltop inflation and natural
inflation. In this class of models,
even-order slow-roll parameters can be order of unity while odd-order slow-roll parameters 
are asymptotically negligible. 
It turns out that it is difficult for $\alpha>-3$ to use it 
to explain the observed Universe due to the anomalous super-Hubble evolution of curvature 
perturbations. That is, for the model parameter $\alpha$ which yields a slightly red-tilted 
power spectrum of curvature perturbations, they grow outside the Hubble radius. In order to 
reproduce the observed amplitude of fluctuations in the presence of such growth, we need 
an extremely low energy scale of inflation, which is much smaller than BBN bound. 
Therefore, the case with $\alpha>-3$ is not observationally feasible.
On the other hand, for the constant-roll model \eqref{Vsol2-} for $\alpha<-3$,
the curvature perturbation has a constant mode and a decaying mode on super-Hubble scales, 
as the standard slow-roll inflation does.
Therefore, the model \eqref{Vsol2-} with $\alpha \lesssim -3$ is 
a novel analytically solvable and observationally viable 
inflationary model with a constant rate of roll, 
which possesses an attractor background evolution, 
slightly red-tilted scalar spectrum, 
and conservation of the curvature perturbation on super-Hubble scales. 
For a realistic model, we have to cut the potential \eqref{Vsol2-} 
somewhere before it becomes negative, 
and it has to be changed after that in order to have subsequent reheating and radiation-dominated regimes.
For the best-fit choice of its 
parameters, this model can reproduce the measured value of the slope of the primordial power
spectrum of scalar (density) perturbations $n_s\approx 0.96$. The prediction for the tensor-to-scalar
ratio $r$ depends on the potential cutoff location, and can be both less than $1\%$, and in the range
$1\% \lesssim r \lesssim 10\%$, too, that is interesting for future search of primordial gravitational waves
from inflation.

\begin{acknowledgments}
The authors thank J.~D.~Barrow and T.~Suyama for useful discussions. 
H.M.\ was partially supported by Japan Society for the Promotion 
of Science (JSPS) Postdoctoral Fellowships for Research Abroad and 
J.Y.\ by JSPS Grant-in-Aid for Scientific Research No.\ 23340058. 
A.S.\ acknowledges RESCEU hospitality as a visiting professor. 
He was also partially supported by the grant RFBR 14-02-00894 and 
by the Scientific Programme ``Astronomy'' of the Russian Academy of Sciences. 
\end{acknowledgments}

\bibliography{refs}

\end{document}